\documentclass[aps,pra,twocolumn,reprint,floatfix, superscriptaddress]{revtex4-2}

\usepackage{subfigure}
\usepackage{psfrag,graphicx}
\usepackage{pict2e}
\usepackage{dcolumn}
\usepackage{amsmath,amssymb,braket}
\usepackage{bm}
\usepackage{color}
\usepackage{latexsym}
\usepackage{epstopdf}
\usepackage{color}
\usepackage[english]{babel}
\UseRawInputEncoding
\usepackage{amsfonts}
\usepackage{bm}
\usepackage{natbib}
\usepackage{bbold}
\usepackage{float}
\usepackage{enumerate}
\definecolor{myblue}{rgb}{0.2,0.2,0.8}
\definecolor{myzard}{cmyk}{0,0,0.05,0}
\definecolor{mywhite}{rgb}{1,1,1}
\definecolor{mywhite}{rgb}{1,1,1}
\definecolor{myred}{rgb}{1,0.,0.3}
\usepackage[colorlinks=true,citecolor=myblue,linkcolor=myred]{hyperref}

\definecolor{mygrey}{gray}{0.35}
\definecolor{myblue}{rgb}{0.2,0.2,0.8}
\definecolor{myzard}{cmyk}{0,0,0.05,0}
\definecolor{mywhite}{rgb}{1,1,1}
\definecolor{mywhite}{rgb}{1,1,1}
\definecolor{myred}{rgb}{1,0.,0.3}

\def\be{\begin{equation}}
\def\ee{\end{equation}}
\def\ba{\begin{align}}
\def\enda{\end{align}}
\def\bi{\begin{itemize}}
\def\ei{\end{itemize}}

\def\beq{\begin{equation}}
\def\beq{\begin{equation}}
\def\eeq{\end{equation}}

\def\kk{{\textbf k}}

\def\rr{{\textbf r}}
\def\KKK{{\textbf K}}

\definecolor{green}{rgb}{0.31, 0.49, 0.16}

\definecolor{purple}{rgb}{0.42, 0.05, 0.68}
\newcommand{\paloma}[1]{{\color{purple}#1}}

\begin{document}

\title{Manipulating Generalized Dirac Cones In Quantum Metasurfaces}

\author{Mar\'ia Blanco de Paz}
\affiliation{Donostia International Physics Center, 20018 Donostia-San Sebastián, Spain}
\affiliation{Instituto de Telecomunica\c c\~oes, Instituto Superior Tecnico-University of Lisbon, Avenida Rovisco Pais 1, Lisboa, 1049‐001 Portugal}

\author{Alejandro Gonz\'alez-Tudela}
\affiliation{Institute of Fundamental Physics IFF-CSIC, Calle Serrano 113b, 28006 Madrid, Spain.}

\author{Paloma A. Huidobro}
\email{p.arroyo-huidobro@lx.it.pt}
\affiliation{Instituto de Telecomunica\c c\~oes, Instituto Superior Tecnico-University of Lisbon, Avenida Rovisco Pais 1, Lisboa, 1049‐001 Portugal}

\begin{abstract}
We discuss the emergence and manipulation of generalised Dirac cones in the subradiant collective modes of quantum metasurfaces. We consider a collection of single quantum emitters arranged in a honeycomb lattice with subwavelength periodicity. While conventional honeycomb quantum metasurfaces host bound modes that display Dirac cones at the $K$ and $K'$ points, we show that introducing uniaxial anisotropy in the lattice results in modified dispersion relations. These include the tilting of Dirac cones, which changes the local density of states at the Dirac point from vanishing (type I) to diverging (types II and III), the emergence of semi-Dirac points, with linear and quadratic dispersions in orthogonal directions, and the anisotropic movement of Dirac cones away from the $K$ and  $K'$ points. Such energy dispersions can modify substantially the dynamics of local probes, such as quantum emitters, to which they have been shown to induce anisotropic, power-law interactions.
\end{abstract}

\maketitle

\section{Introduction}
Metasurfaces enable the tuning of light-matter interactions by exploiting collections of subwavelength nano-antennas made of metallic or dielectric nanostructures, even realising properties that are not available in natural materials~\cite{smith2004metamaterials,yu2014flat,meinzer2014plasmonic}. Conventionally, metasurfaces have been designed as periodic arrays of nanoresonators such as as dielectric or plasmonic nanoantennas. These setups have successfully led to multiple applications, including sensing \cite{kravets2018plasmonic} and nanoscale lasing \cite{wang2018rich}, as well as topological protection \cite{Poddubny2014Topological,yves2017crystalline,Pocock2018,downing2019topological,proctor2019exciting}, polaritonic edge states \cite{DowningM2021Polaritonic} and even gauge fields in subwavelength arrays \cite{mann2020tunable}.

Recently, the possibility of realising quantum metasurfaces is being considered~\cite{zhouOptical2017,Bekenstein2020,solntsev2021metasurfaces}. In their most fundamental realisation, these make use of single photon emitters such as atomic transitions as the most elementary subwavelength antennas ~\cite{zumofen2008perfect}, that operate at the single photon level and with very low radiative loss \cite{bloch2005ultracold}. 
By arranging single photon emitters in lattices of subwavelength periodicities, coherent dipole-dipole interactions between all the quantum emitters in the lattice result in cooperative effects that lead to drastic changes in the optical properties of the emitters when they are placed in the array \cite{jenkins2012controlled,olmos2013longrange,bettles2016enhanced,schilder2016polaritonic,ruostekoski2016emergence,shahmoon2017cooperative,zhou2017optical}. An important instance of cooperative effects are the sub-radiant optical states whose coupling to the photonic environment is greatly reduced. Sub-radiant modes can be harnessed for selectively improving radiation in a given desired channel~\cite{asenjogarcia17a}, for mediating non-trivial emitter-emitter interactions when additional atoms are placed nearby~\cite{Masson2020,Patti2021,Brechtelsbauer2020,Castells-Graells2021AtomicDimers,Fernandez-Fernandez2021TunableMetasurfaces}, generating topological edge modes~\cite{perczel17a,bettles17a} or creating magnetic responses at optical frequencies~\cite{alaee2020quantum,ballantine2020optical}.  These exciting perspectives have triggered experimental interest in the topic, which has already crystallized in the first experimental realization of a sub-radiant optical mirror~ \cite{rui2020subradiant}.
Among the prospects of quantum metasurfaces for controlling the interaction between photons and optical media is the generation of non-trivial photonic energy dispersions, like the ones one can obtain with photonic crystals~\cite{Chang2018}, that lead to extreme and exotic forms light-matter interactions and their exploitation as quantum photonic setups~\cite{Masson2020,Patti2021,Brechtelsbauer2020,Castells-Graells2021AtomicDimers,Fernandez-Fernandez2021TunableMetasurfaces}.

\begin{figure}[hb]
   \centering
  \includegraphics[width=0.9\columnwidth]{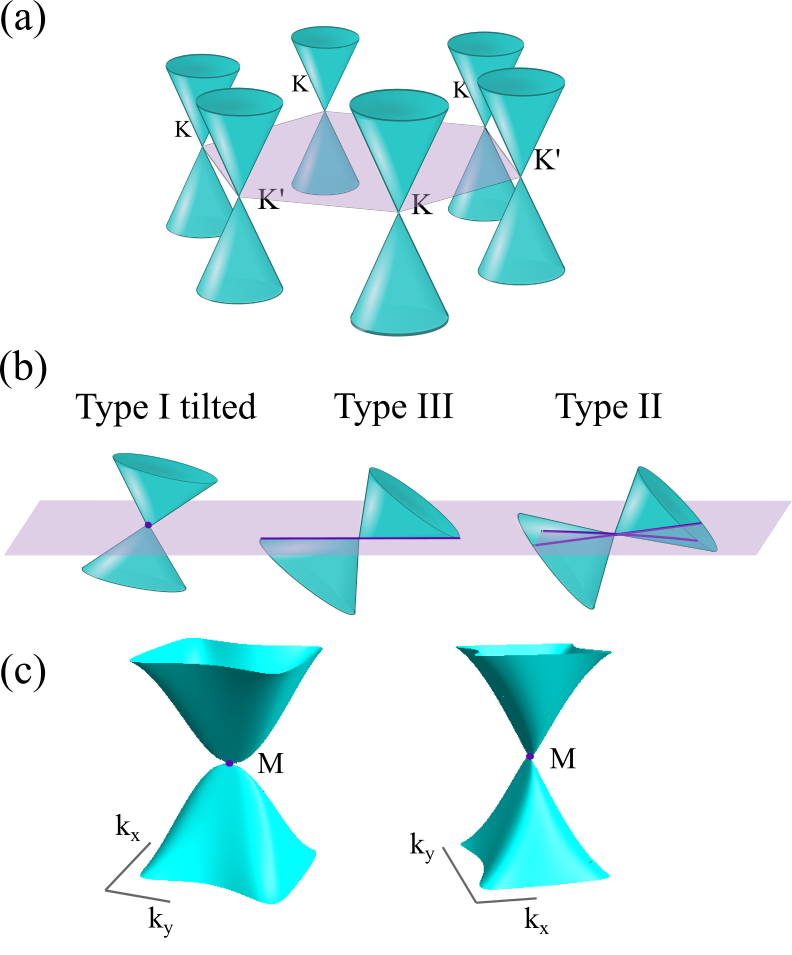}
   \caption{Dirac and generalised Dirac dispersions. (a) Sketch of the Dirac cones of the conventional honeycomb lattice distributed in the Brillouin zone. Introducing anisotropy in the lattice allows to manipulate the crossings and generate tilted cones (b) of types I (left), II (right), and III or critical (middle), as well as semi-Dirac cones (c), which present linear and quadratic dispersions in orthogonal directions in reciprocal space.}
   \label{fig0}
\end{figure}

On the other hand, since the discovery of the remarkable electronic properties of graphene and how these emerge from isotropic linear energy dispersions~\cite{castroneto09a}, the realisation of synthetic Dirac dispersions has attracted much attention in photonics, as a variety of photonic platforms enable the possibility of synthetising and even manipulating a plethora of unconventional energy dispersions \cite{Haldane2008Possible,Sepkhanov2007Extremal,Zandbergen2010Experimental,huang2011Dirac,bravo2012Enabling,yang2018ideal,Hu2018Type,mann2018,milicevic2019typeiii,Real2020SemiDirac,Kim2020Universal}. Conventional Dirac cones such as the ones found in graphene present an isotropic and linear energy dispersion and are extremely robust against perturbations of the lattice. The reason for this is that they have a topological origin: in the presence of time-reversal and inversion symmetry, they appear in pairs of opposite topological charge and can only disappear by merging with a cone of the opposite charge \cite{Montambaux2009Merging,tarruell2012creating,Bellec2013Topological}. On the other hand, generalised Dirac dispersions are possible. Figure \ref{fig0} schematically shows the conventional dispersions in Dirac media (a), as well as the new generalised dispersions enabled by anisotropy. First, Dirac cones can be tilted (b), such that the isofrequency contours at the degeneracy are either still a point (type I, depicted left), a line (critical, or type III, middle) or two lines (type II, right) \cite{Goerbig2008Tilted}. This involves remarkable changes in the density of states, which changes from zero to infinity. On the other hand, anisotropic cones where the dispersion is linear along one direction but becomes quadratic along the orthogonal one are also possible, as sketched in panel (c), and named semi-Dirac cones \cite{Hasegawa2006Zero,Dietl2008NewMagnetic}. 

These unconventional structured photonic baths have the potential of greatly modifying the quantum dynamics of probe emitters placed in the vicinity of the metasurface. Dirac cones have already been shown to lead to exotic quantum dynamics and long-range photon-mediated interactions~\cite{Gonzalez-Tudela2018,Perczel2020a,Navarro-Baron2021Photon-mediatedSlab,Perczel2020b}. A quantum emitter tuned to a Dirac point displays an unconventional (non-exponential) decay despite the vanishing density of states, which is attributed to a power-law photonic mode localised around the emitter. This so-called quasi-bound state can mediate long range interactions between quantum emitters, which can also be made anisotropic by employing generalised Dirac dispersions~\cite{redondoyuste2021quantum}.


In this work we present a quantum metasurface that hosts generalised Dirac dispersions, and discuss how they emerge and how to manipulate them. We do so by introducing uniaxial anisotropy in a subwavelength honeycomb lattice of single photon quantum emitters. The lattice anisotropy enables the emergence of semi-Dirac and tilted Dirac cones. The paper is structured as follows. We start by outlining the theoretical framework that allows us to calculate the dispersion relations in these subwavelength lattices with long range interactions. After revisiting the properties of Dirac cones in honeycomb quantum metasurfaces, we first consider the out-of-plane modes of anisotropic honeycomb quantum metasurfaces and demonstrate the emergence of semi-Dirac cones and the anisotropic displacement of Dirac cones in reciprocal space. Next, we consider the in-plane modes where we find another instance of semi-Dirac cones, as well as tilted Dirac cones and anisotropic displacements of Dirac cones. Finally, we present a more detailed discussion on the effect of retardation in these generalised dispersions.  


\begin{figure}[tb!]
  \centering
  \includegraphics[width=0.7\columnwidth]{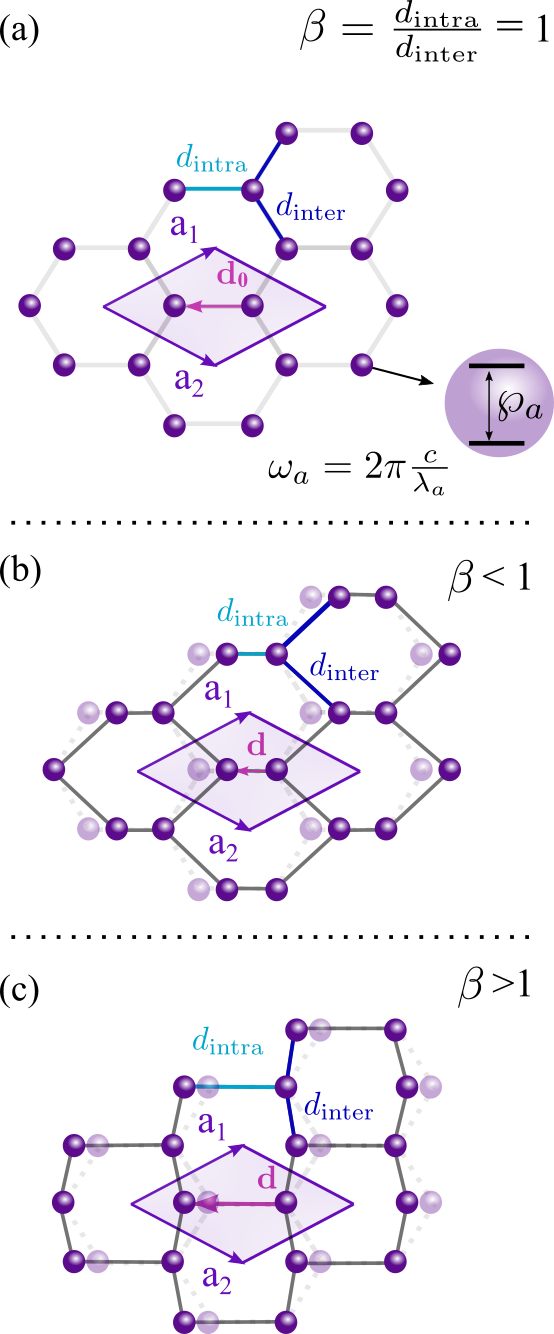}
  \caption{Geometry of the quantum metasurfaces: a periodic array of single quantum emitters arranged in a non-Bravais lattice with two sites per unit cell. Starting from a honeycomb lattice (a), anisotropic lattices are generated by pushing together (b) or apart (c) along the horizontal direction the two emitters contained in the unit cell. In all cases the basis vectors $\mathbf{a}_{1/2}$ and the unit cell is kept the same, and what varies is the ratio between the intra- and intercell nearest neighbour distances, $d_\mathrm{intra}$ and $d_\mathrm{inter}$, which characterises the degree of anisotropy through the parameter $\beta=d_\mathrm{intra}/d_\mathrm{inter}$.  }
  \label{fig1}
\end{figure}

\section{Anisotropic lattices and theoretical framework} 
We consider a generalised version of the honeycomb lattice that allows us to realise anisotropic arrays. As depicted in Fig.~\ref{fig1}, these are non-Bravais lattices with two atoms per unit cell, described by primitive vectors $\mathbf{a}_{1/2}=d_0\sqrt{3}/2(\sqrt{3}\hat{\bf{e}}_x\pm \hat{\bf{e}}_y)$, with $d_0$ being the nearest neighbour distance of the honeycomb lattice [panel (a)], and with basis vector $\mathbf{d}=-d_\mathrm{intra}\hat{\mathbf{e}}_x$. Here, $d_\mathrm{intra}$ and, correspondingly, $d_\mathrm{inter}$, represent the intracell and intercell nearest neighbour distances. In order to realise anisotropic lattices, we allow for these two distances to be different, $d_\mathrm{intra}\neq d_\mathrm{inter}$, and their ratio defines the anisotropy parameter, $\beta =d_\mathrm{intra}/d_\mathrm{inter}$ \paloma{\footnote{We note that the maximum value $\beta$ can take is $\beta_{\mathrm{max}}=1.7321$. For this value of $\beta$, the anisotropy is maximum, corresponding to a rectangular lattice, and values of 
$\beta>\beta_{\mathrm{max}}$ can be mapped to $\beta<\beta_{\mathrm{max}}$. On the other hand, we keep $\beta\gtrsim0.5$ to avoid unphysically close distances between the emitters.}}. 
For a standard honeycomb lattice, $d_\mathrm{intra}=d_\mathrm{inter}=d_0$, such that the distance between nearest neighbours is the same in all directions, and $\beta=1$. On the other hand, $\beta<1$ represents an anisotropic lattice where the two atoms in the unit cell are pushed together [see Fig.~\ref{fig1} (b)], which creates an unbalance in the nearest neighbour distance along different directions, $d_\mathrm{intra}\neq d_\mathrm{inter}$. Similarly, for $\beta>1$ the two atoms in the unit cell are pushed apart. The unbalanced nearest neighbour distances affects the interactions between all elements in the lattice and results in a variety of generalised Dirac cones, as we discuss in detail in the following. 
Finally, while $d_0$ is the nearest neighbour distance only in the honeycomb case, we will use it to characterise the periodicity of all the lattices since it defines the size of the lattice vectors as $|\mathbf{a}_{1/2}|=2\sqrt{3}d_0$.

Each lattice site in the 2D array contains a single quantum emitter that we model as a two level system with resonance frequency $\omega_a=2\pi c/\lambda_a$ and polarization dipole $\boldsymbol{\wp_a}$. The dynamics of the quantum emitters array can be described through an effective non-Hermitian Hamiltonian~\cite{asenjogarcia17a,Perczel2017}:
\begin{align}
    \frac{H}{\hbar} = \sum_{j=1}^{N_A}\left(\omega_a-i\frac{\Gamma_a}{2}\right) \sigma_{ee}^j+  \sum_{\substack{j=1 \\ i\neq j}}^{N_A}\left(J_{ij}-i\frac{\Gamma_{ij}}{2}\right)\sigma_{eg}^i\sigma_{ge}^j\,,
\label{eq:H_eff}
\end{align}
where $j$ is a index running over all emitters in the metasurface ($N_A$), placed at positions $\rr_j$, and $\Gamma_a=|\boldsymbol{\wp_a}|^2\omega_a^3/(3\pi\hbar c^3)$ is the individual free-space decay rate. The coherent ($J_{ij}$) and incoherent ($\Gamma_{ij}$) photon-mediated interactions among emitters are given by the free space Green's dyadic $\mathbf{G}_0(\rr_i-\rr_j)$~\cite{lehmberg70a,lehmberg70b}:
\begin{eqnarray}
    J_{ij}=-\frac{3\pi \Gamma_a c}{\omega_a}\textrm{Re}\left[\hat{\boldsymbol{\wp}}^*_{i} \cdot\mathbf{G}_0(\rr_i-\rr_j)\cdot\hat{\boldsymbol{\wp}}_{j} \right]
\label{eq:J} \\ 
    \frac{\Gamma_{ij}}{2}=\frac{3\pi \Gamma_a c}{\omega_a}\textrm{Im}\left[\hat{\boldsymbol{\wp}}^*_{i} \cdot\mathbf{G}_0(\rr_i-\rr_j)\cdot\hat{\boldsymbol{\wp}}_{j}\right]
\label{eq:G}
\end{eqnarray}
where $\hat{\boldsymbol{\wp}}_i=\boldsymbol{\wp}_{i}/|\boldsymbol{\wp}_i|$,
\begin{equation}
\begin{split}
    \mathbf{G}_0(\rr)= \frac{1}{4\pi}\left[\mathbb{1}+\frac{\nabla\otimes\nabla}{k_0^2}\right]\frac{e^{ik_0 |\rr|}}{|\rr|}\,,
\end{split}
\label{eq:Green_tensor1}
\end{equation} 
and $k_0=\omega/c$. Importantly, the dipole-dipole interactions described by the Green's function are long-ranged, with terms that decay as $1/r^3$, $1/r^2$ and $1/r$. Additionally, interactions involve all the elements in the lattice, and depend on the polarization of the electromagnetic fields. The geometry of the array generates two sets of modes: in-plane modes where the dipole's polarisation is contained in the plane of the array, and out-of-plane modes with dipoles polarised orthogonal to the lattice plane. 

The eigenstates of the quantum metasurface described by the above Hamiltonian can be found in the single excitation subspace by looking for Bloch modes, 
\begin{equation}
 S^\dagger_\kk=\frac{1}{\sqrt{N}}\sum_{n=1}^N\sum_{m=1}^2 \sigma^{n,m}_{eg} e^{i\kk\cdot\rr_n}, 
\end{equation}  
where $\kk=(k_x,k_y)$ is the mode wavevector in the plane of the array and we take into account that we have a non-Bravais lattice, with the sums now running over $n$, up to the total number of unit cells ($N$) and over the two sites per unit cell ($m$). Obtaining the eigen-energies of the Bloch modes from the above Hamiltonian then reduces to diagonalising the following matrix~\cite{Perczel2017}:
\begin{align} \label{eq:Mk}
    \mathbf{M}^{\alpha\beta,\mu\nu}_\kk&=(\omega_a-i\frac{\Gamma_a}{2})\delta_{\alpha\beta}\delta_{\mu\nu} \\ &
    - \frac{3\pi \Gamma_a c}{\omega_a} \left[ \sum_{m=1}^2 \sum_{\bf{R}_n\neq0} e^{-i\kk\cdot\bf{R}_n} G^{\alpha\beta}_0(\bf{R}_n)\delta_{m\mu}\delta_{m\nu} \right. \nonumber \\ &
    + \sum_{\bf{R}_n} e^{-i\kk\cdot\bf{R}_n} G^{\alpha\beta}_0(\bf{R}_n+\mathbf{d})\delta_{1\mu}\delta_{2\nu} \nonumber \\ &
    +\left. \sum_{\bf{R}_n} e^{-i\kk\cdot\bf{R}_n} G^{\alpha\beta}_0(\bf{R}_n-\mathbf{d})\delta_{2\mu}\delta_{1\nu} \right] \nonumber,
\end{align}
where $\{\bf{R}_n\}$ represent the in-plane position vectors of all the unit cells in the lattice. Index $n$ runs from 0 to $N$, $\{\alpha,\beta\}$ run over the 3 spatial degrees of freedom, and $\{\mu,\nu\}$ over the 2 lattice sites. Hence $ \mathbf{M}_\kk$ is a 
$6\times6$ matrix with eigenvalues $\omega_\kk-i\frac{\gamma_\kk}{2}$, which represent the photonic band structure ($\omega_\kk$) and the radiative decay of the Bloch modes ($\gamma_\kk$). The lattice sums of the Green's tensor that appear in Eq.~\ref{eq:Mk} are slowly convergent due to the long range interactions and we employ the Ewald method to perform them efficiently~\cite{ewald1921,Linton2010}.

\section{Dirac cones in a honeycomb quantum metasurface} 

\begin{figure}[tb]
   \centering
  \includegraphics[width=0.99\columnwidth]{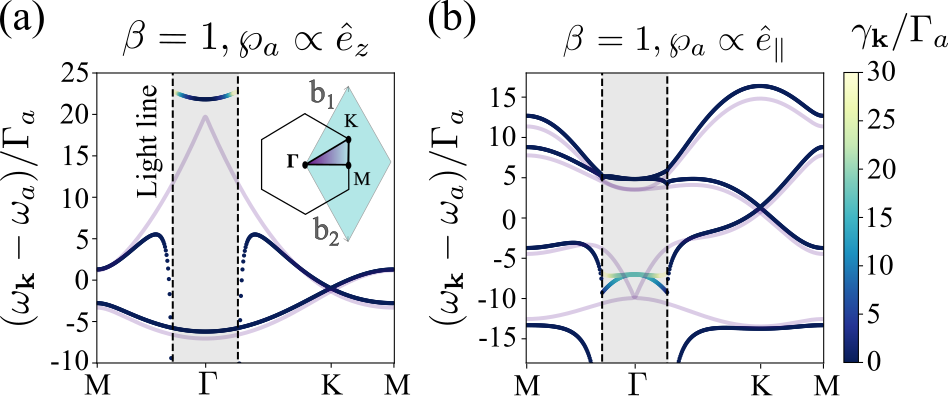}
   \caption{Photonic band structure of honeycomb quantum metasurfaces. (a,b) Dispersion relations of out-of-plane (a) and in-plane (b) modes for a honeycomb lattice of single quantum emitters with resonance frequency $\omega_a$ and individual decay rate $\Gamma_a$. Dirac crossings are visible at the $K^{(')}$ points for both sets of modes. The decay rate of the modes is color coded, with the scale bar given in (b). Results from a quasistatic approximation are plotted as a light grey line. The region within the light cone is shaded in gray. The periodicity of the lattice is determined by the nearest neighbour distance, $d_0=0.1\lambda_a$. }
   \label{fig2}
\end{figure}


We first apply the theoretical formalism to revisit the case of a honeycomb quantum metasurface \cite{Perczel2017,bettles17a}. Figure \ref{fig2} presents the photonic band structure for the two out-of-plane modes, $ \hat{\boldsymbol{\wp}}_i \propto \hat{e}_z$ (a), and the four in-plane modes, $ \hat{\boldsymbol{\wp}}_i \propto \hat{e}_{\|}$ (b), of the honeycomb lattice, $\beta=1$. The array periodicity is given by the chosen nearest neighbour distance, $d_0=0.1\lambda_a$, with $\lambda_a=2\pi c/\omega_a$, here and throughout this work unless otherwise stated. This distance determines the position of the light line, which separates radiative and non-radiative modes. In the region within the light line, shaded in gray, modes have a non-zero radiative width, which is color coded in the plot [scale bar in panel (b)]. As seen in the figure, the most radiative modes present strong interactions with the light line, and beyond the light cone modes are non-radiative. It is in the region outside the light cone that we can observe Dirac points for both sets of modes, appearing at the $K^{(')}$ points as sketched in (c). Additionally, we also plot the band structure resulting from a quasistatic approximation, where we neglect the medium and long range terms in the Green's function and keep only the shortest range one ($1/r^3$), but we still sum over all the lattice sites. This approximation is valid for very subwavelength arrays, as the short period quantum emitter arrays considered here, but since it does not include retardation it fails to predict the correct interactions with the light line for the most radiative modes, as can be seen in the figure. The fact that interactions among all the lattice elements are included results in the breaking of the chiral symmetry, which is characteristic of the honeycomb lattice in the nearest neighbour limit, and explains the lack of symmetry of the bands with respect to $\omega_\kk = \omega_a$. However, for these set of parameters the Dirac cones still appear close to $\omega_{\KKK} = \omega_a$. 
%
%
%


\section{Generalised Dirac cones in the out of plane modes of anisotropic quantum metasurfaces}

In this Section we discuss the emergence and evolution of generalised Dirac cones for the out of plane modes of anisotropic quantum metasurfaces. The anisotropic lattices are shown in Figure~\ref{fig1}(b) and (c) for $\beta>1$ and $<1$, respectively. As described above, the lattice vectors of the standard honeycomb, and hence the lattice periodicity, are kept, while the basis vector is changed away from its honeycomb value, pushing together/apart the two sites in the unit cell. As we will see, the anisotropy of the interactions in the new lattices enables the emergence of generalised Dirac dispersions. 

\begin{figure}[tb!]
  \centering
  \includegraphics[width=1\columnwidth]{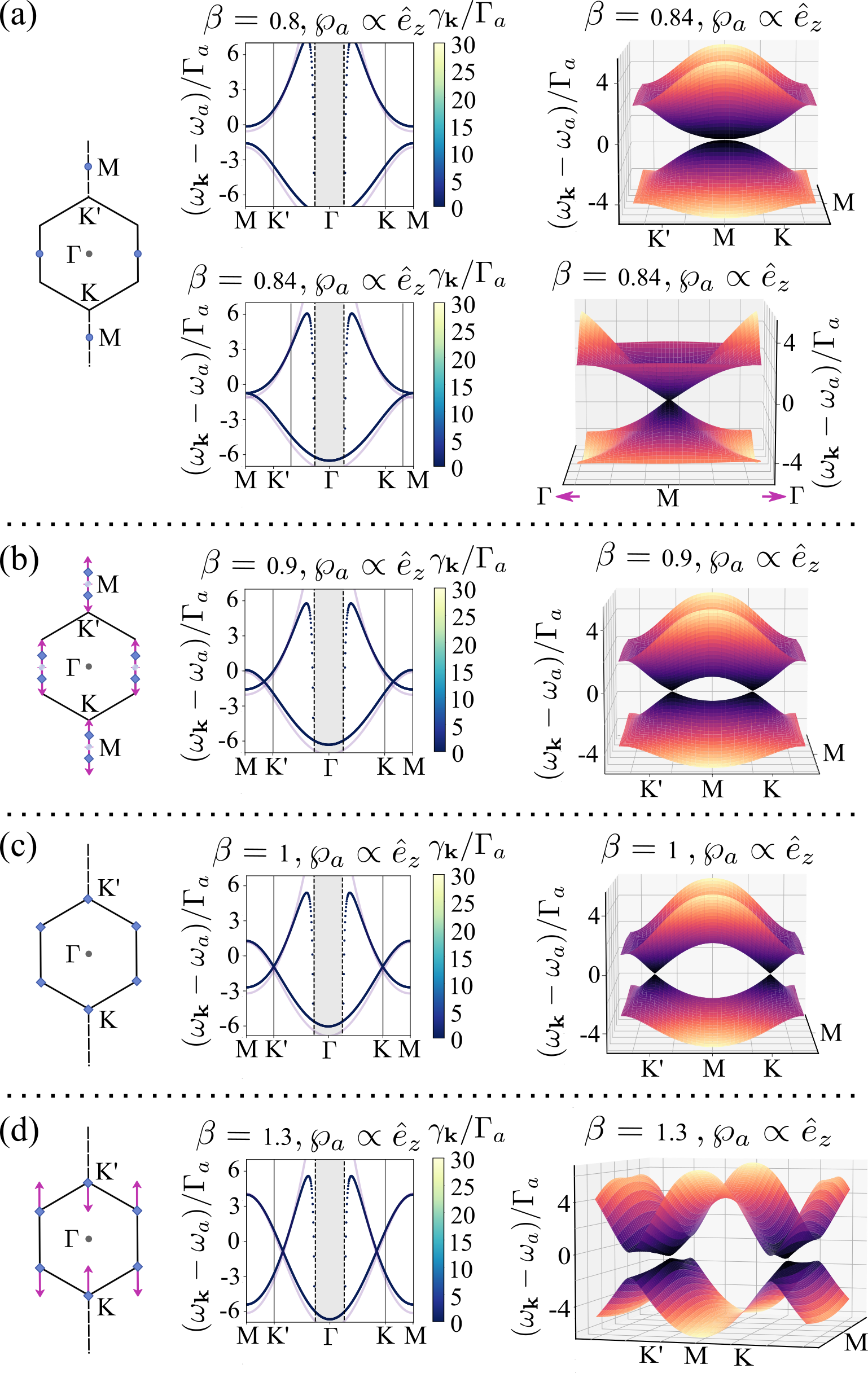}
  \caption{Generalised Dirac cones in the out of plane modes of anisotropic honeycomb lattices. (a) Semi-Dirac cones emerge at the $M$ points for a critical value of anisotropy, $\beta_c=0.84$. Dispersion is quadratic along the $KMK'$ direction (shown in the line and right 3D plot), while it is linear in the orthogonal direction, $\Gamma M \Gamma$ (see left 3D plot).
  (b) As anisotropy decreases and the geometry of the honeycomb lattice is approached, $\beta_c<\beta<1$, the semi-Dirac cones split into two Dirac cones that move away from $M$ and towards $K$ and $K'$. This is shown for $\beta=0.9$. 
  (c) When $\beta=1$ the Dirac cones are at the $K$ and $K'$ points, as corresponds to a honeycomb lattice. 
  (d) For $\beta>1$, an anisotropic distribution of Dirac cones in reciprocal space develops. All Dirac points move vertically in the $K' \Gamma$ direction. In all cases the periodicity of the lattice is fixed by $d_0 = 0.1\lambda_a$. In the line plots, the quasistatic band structure is shown also for comparison as a light gray line. }
  \label{fig3}
\end{figure}

Figure \ref{fig3} presents the evolution of the band structures for quantum metasurfaces with the same lattice constant as in figure \ref{fig2}, fixed by $d_0=0.1\lambda_a $, as anisotropy changes from a value $\beta<1$ (a,b), through $\beta=1$ (c) and to $\beta>1$ (d). In all panels, a sketch of the position of the generalised Dirac points in reciprocal space is shown in the left insets. The middle insets present line plots of the photonic dispersion along a vertical path path in reciprocal space between two $M$ points, such that it passes through the $K'$, $\Gamma$ and $K$ points. Additionally, the right insets display 3D plots of the generalised cones. 

We start with a critical value of anisotropy, $\beta_c=0.84$, for which the two out of plane modes cross in a semi-Dirac cone at the $M$ point. As seen in Fig.~\ref{fig3}(a), the dispersion is quadratic along the $KMK'$ direction, and linear in the orthogonal direction, $\Gamma M \Gamma$. These kind of generalised Dirac dispersions have been shown to yield long range anisotropic interactions with quantum emitters placed close to the metasurface~\cite{redondoyuste2021quantum}. Interestingly, these denegeracies have a zero topological charge~\cite{milicevic2019typeiii}. Indeed, for higher anisotropy, $\beta<\beta_c$, the bands are gapped, while for lower anisotropy, $\beta>\beta_c$, the semi-Dirac cones split up into pairs of Dirac cones of opposite topological charge. That is, increasing $\beta$ towards $\beta=1$ splits up the semi-Dirac cones into pairs of Dirac cones that move along the vertical direction in $k-$space towards the $K$ and $K'$ points. The pair of Dirac cones are shown in panel (b) for $\beta=0.9$, and they display an anisotropic spatial distribution in reciprocal space, since they move in the vertical direction. Next, the cones reach the $K$ and $K'$ points for $\beta=1$, as corresponds to a honeycomb lattice, shown in panel (c). By increasing anisotropy again with $\beta>1$, the Dirac cones continue moving vertically in reciprocal space past the $K^{(')}$ points and towards the $\Gamma$ point, as shown in panel (d) for $\beta=1.3$. This results in a compression of the Dirac cones distribution in reciprocal space along the vertical direction ($k_y$), as well as in a flattening of the bands along the $k_x$ direction. Hence, the cones are highly anisotropic, with a much lower slope in the $k_x$ than in the $k_y$ directions, as can be seen in the 3D plot in panel (d). As anisotropy is increased towards the maximum value ($\beta=1.7321$), the anisotropy of the cones increases further with the slope along the $k_x$ direction away from the crossings approaching zero. Additionally, the position of the cones moves further, with the three top (and the three bottom) cones sketched the figure tending to align with each other in the vertical direction.  


\section{Generalised Dirac cones in the in plane modes of anisotropic quantum metasurfaces}

\begin{figure}[tb!]
  \centering
  \includegraphics[width=0.95\columnwidth]{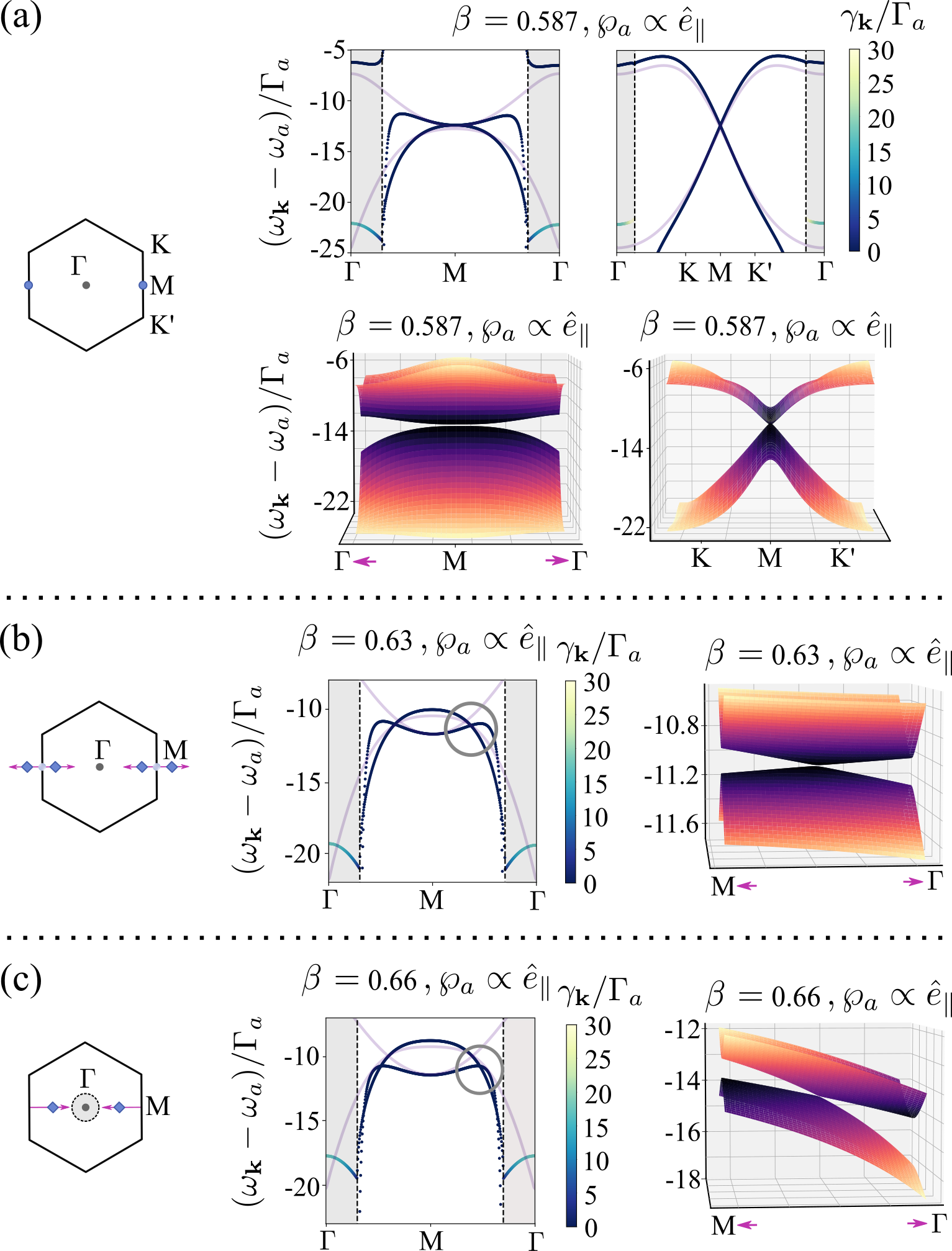}
  \caption{Semi-Dirac and tilted Dirac cones for the in-plane modes of anisotropic honeycomb quantum metasurfaces. (a) Semi-Dirac point formed by the crossing of the lowest two bands for a critical value of anisotropy, $\beta=0.587$. Dispersion along the $\Gamma M \Gamma$ and $ KMK'$ directions is shown in the left and right line and 3D plots, respectively.  (b,c) As anisotropy decreases, the semi-Dirac points split into two Dirac points that travel in the $k_x$ direction towards $\Gamma$. At the same time, interactions with the light line tilt the cones. For $\beta=0.63$ the tilted cone is type I tilted (b), while for $\beta=0.66$ the cones are closer to the light line and become type II (c). The periodicity of the lattice is fixed as $d_0 = 0.1\lambda_a$. In the line plots, the quasistatic band structure is shown also for comparison as a light gray line.  }
  \label{fig4}
\end{figure}

After discussing the effect of lattice anisotropy in the dispersion of out of plane modes, we now consider the in plane modes. First, we focus on the lowest two bands of the honeycomb metasurface, see Fig.~\ref{fig2} (b). For these bands a semi-Dirac cone emerges at the $M$ points for a critical value of anisotropy $\beta<1$, which splits into two Dirac cones that move anisotropically in reciprocal space differently than in the case discussed above for the out of plane modes.   

The band structure for the in-plane modes of the metasurface is shown in Fig.~\ref{fig4}(a) for a critical value of anisotropy, $\beta_c=0.587$. As for the out-of plane modes, we see that these two bands touch with quadratic dispersion along one direction and linear in the orthogonal direction. However, differently from the case discussed above, in this case dispersion is quadratic in the horizontal direction ($\Gamma M \Gamma$  path, insets in left column), and linear in the vertical direction ($KMK'$ path, insets in right column). When moving away from this critical value of anisotropy by increasing $\beta$, each semi-Dirac cone splits into two Dirac cones that travel in reciprocal space. In contrast to the behaviour of out-of-plane modes, in this case the Dirac cones travel towards $\Gamma$ along the horizontal directions. This is shown in Fig. \ref{fig4}(b) and (c) as we now describe in detail.

Due to the retarded interactions characteristic of these quantum metasurfaces, as $\beta $ increases towards 1 and the cones travel towards $\Gamma$, they encounter the light line and the bands are strongly affected by interaction with it. This can be seen by comparing the full electrodynamic results (color coded) with the quasistatic approximation (light grey line) in the line plots in panels (b) and (c). In the quasistatic case as $\beta$ increases and the degeneracies move away from $M$, both bands retain their quadratic curvature in the horizontal direction but move in energy such that the quadratic touching point at $M$ transforms into two isotropic type I Dirac points. On the other hand, when fully retarded interactions are taken into account, the top band presents a polaritonic-type splitting at the light line as it is prevented from entering into light cone and is instead strongly bent downwards. This results in the Dirac crossings being tilted. Initially, the tilted cone is type I, as can be seen in the line and 3D plots in panel (b)  for $\beta=0.63$, but as $\beta$ increases more the cone becomes more and more tilted, crossing the critical point where one of the bands is flat at the degeneracy (type III), and becoming type II, as shown for $\beta=0.66$ in panel (c). These type of tilted Dirac cones arising due to strong polaritonic type interactions in subwavelength arrays were also observed in Ref.~\cite{mann2018}, where conventional subwavelength honeycomb lattices embedded inside a cavity where considered. While here they appear by modifying the lattice, in that work they require the encapsulation of the array within two mirrors and tuning its distance. Finally, for a larger increase of $\beta$ the degeneracy between the two bands is lost due to the strong interaction with the light line. In contrast, we note that in the quasistatic approximation the degeneracy is mantained until it reaches the $\Gamma$ point for $\beta=1$, as shown for the honeycomb lattice in Fig. \ref{fig2}(b). This stresses the importance of including fully retarded interactions when studying subwavelength arrays.

\begin{figure}[tb!]
  \centering
  \includegraphics[width=1\columnwidth]{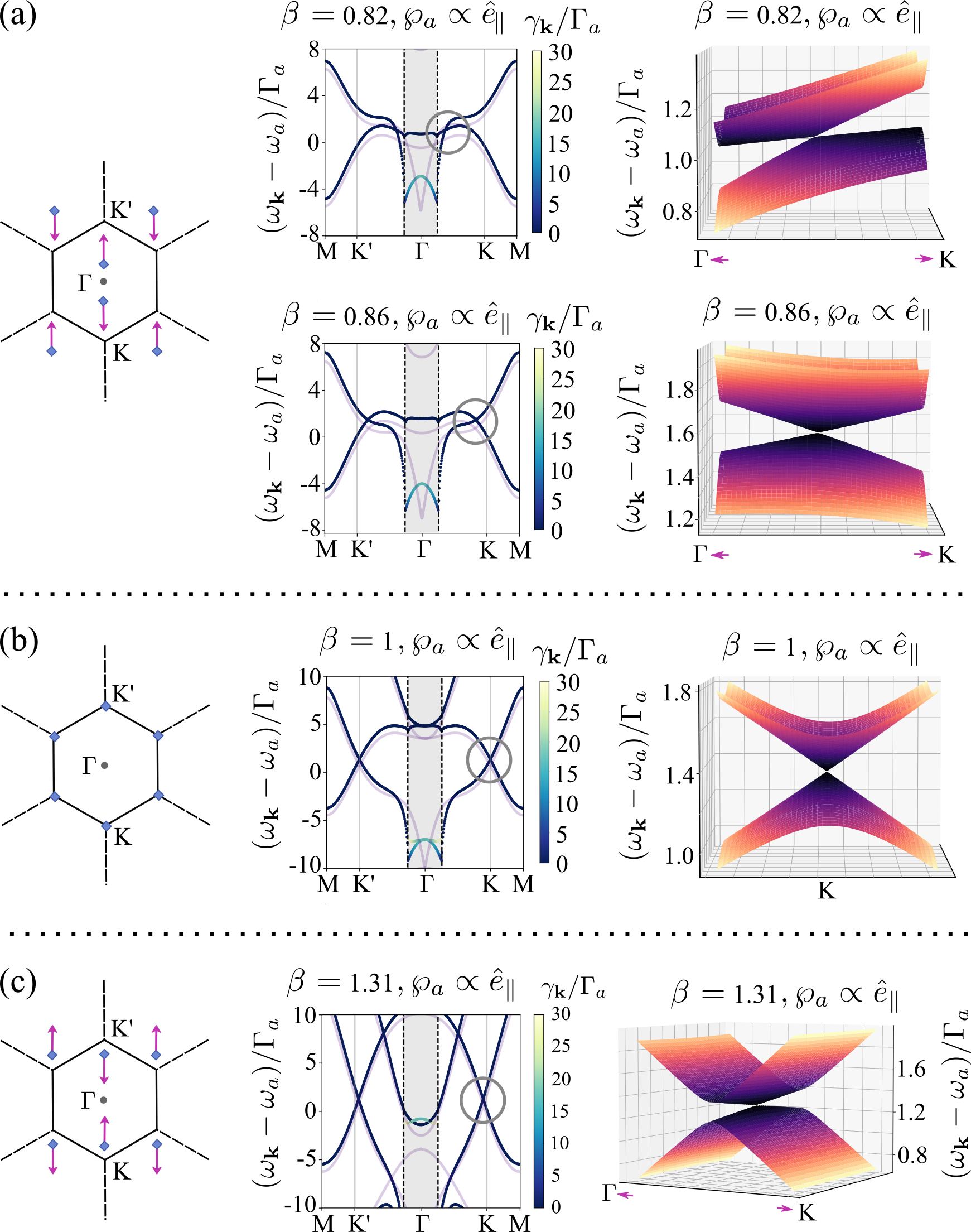}
  \caption{Tilted Dirac cones and anisotropic distribution of Dirac cones in reciprocal space for the two middle in-plane modes of an anisotropic honeycomb metasurface. (a) Tilted Dirac cones appear in the $\Gamma K$ lines for $\beta<1$. For large anisotropy,  $\beta=0.82$ (insets in top row), the cones are type II. As $\beta$ increases the cones transform into type I tilted cones as they approach the $K'$ points ($\beta=0.86$, insets in bottom row). (b) For $\beta=1$ isotropic Dirac cones at the  $K'$ points are found. (c) The cones move anisotropically in reciprocal space for $\beta>1$, and the bands flatten in the horizontal direction. The periodicity of the lattice is fixed as $d_0 = 0.1\lambda_a$. In the line plots, the quasistatic band structure is shown also for comparison as a light gray line.}
  \label{fig5}
\end{figure}

Now, we turn our attention to the two middle bands, and we study the emergence of generalised Dirac cones and their evolution from $\beta<1$ to $\beta>1$ in Fig.~\ref{fig5}. First, for $\beta<1$ we find Dirac cones along the $\Gamma K^{(')}$ lines, travelling away from $\Gamma $ and towards $K^{(')}$ as $\beta$ approaches 1. This is shown in panel (a). For a given value of anisotropy, $\beta=0.82$ (plots in insets in top row), we find that the Dirac cones are tilted and of type-II. For these parameters, they appear very close to the light line. However, and differently from the case discussed above or the work in Ref.~\cite{mann2018}, the type-II tilt of these bands is not created by the interaction between the bands and the light line, since it is also present in the quasistatic bands, as can be seen in the light gray line in the plot. Next, as $\beta$ increases, the cones move along vertical lines towards the $K^{(')}$ points, and the tilt transforms into a type-I tilt, as shown for $\beta=0.86$ (plots in insets in bottom row). Then, as expected, when $\beta=1$ is reached, we observe conventional Dirac cones at the $K^{(')}$ points in panel (b). Increasing anisotropy away from the honeycomb case with $\beta>1$ results in a situation similar to that of the out-of-plane modes described above. As shown in panel (c), the Dirac cones move vertically in reciprocal space away from the $K^{(')}$ points and as they do so they develop a strong anisotropy, with much flatter slopes along $k_x$ than along $k_y$, as is clear from the 3D plot shown in (c).

\section{Retardation effects}
\begin{figure}[ht!]
  \centering
  \includegraphics[width=0.9\columnwidth]{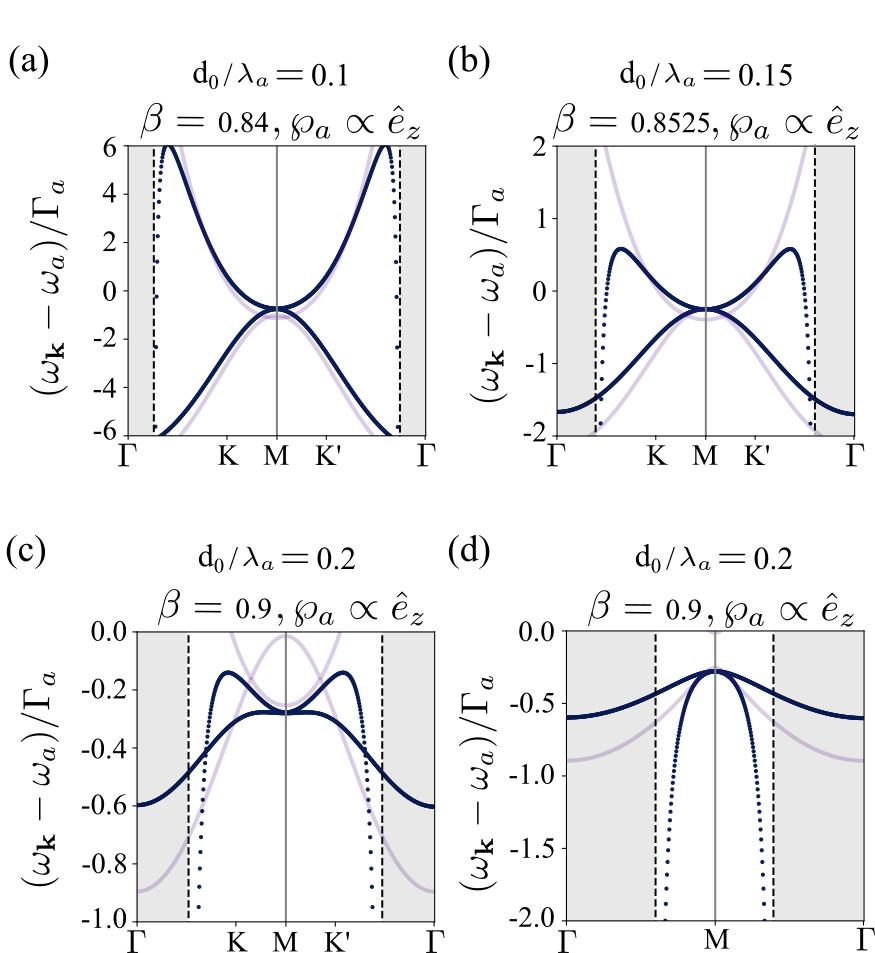}
  \caption{Retardation effects are visible for increasing lattice periodicity. (a-c) Band structures of out of plane modes along the vertical direction in reciprocal space for $d_0/\lambda_a=0.1$, 0.15 and 0.2, and $\beta=0.84$, 0.8525 and 0.9, respectively. In (a) and (b) the degeneracy is a semi-Dirac cone (the linear dispersion in the orthogonal direction is not shown here). In (c-d) the semi-Dirac cone is lost as dispersion is quadratic in both directions due to strong interactions with the light line. }
  \label{fig6}
\end{figure}

Finally, in this Section we further discuss the effect of retardation by considering lattices of increasing periodicity. We focus on the semi-Dirac cones that emerge for out-of-plane modes. Figure \ref{fig6}(a-c) presents band structures along a vertical path in reciprocal space between $\Gamma$ and $M$ for increasing values of $d_0/\lambda_a=0.1$, 0.15 and 0.2. We recall that $d_0$ gives the nearest neighbour distance of the corresponding honeycomb lattice ($\beta=1$), and it fixes the length of the lattice vectors as $|\mathbf{a}_{1, 2}|=2\sqrt{3}d_0\approx 0.35\lambda_a,\, 0.52\lambda_a,$ and $0.7\lambda_a$, respectively for each case. 
First, in panel (a) we reproduce the quadratic dispersion featured by the semi-Dirac cone for $d_0/\lambda_a=0.1$ and $\beta=0.84$ discussed already in Fig.~\ref{fig3}(a). In panel (b) we increase the lattice periodicity to $d_0/\lambda_a=0.15$, and we see how the light cone moves further away from $\Gamma$, as expected. Additionally, the polariton-type interaction between the top band and the light line becomes stronger. However, there is still a semi-Dirac point for a larger value of anisotropy than in the previous case, $\beta=0.8525 $, with quadratic dispersion along the vertical direction as shown in panel (b), and linear dispersion along the orthogonal direction (not shown here). We stress that since the semi-Dirac points emerge at critical values of anisotropy and the bands depend on the lattice periodicity, the anisotropy value where the semi-Dirac cones emerge depends on the lattice periodicity. Finally, we increase the periodicity further to $d_0/\lambda_a=0.2$ in panel (c), where retarded interactions become even more important and the bands are more strongly affected. For this case, we can still find a quadratic touching point for $\beta=0.9$. However, the linear crossing in the orthogonal direction is lost due to the strong modification of the bands owing to retardation, and the degeneracy is quadratic also in this direction, see panel (d).  Interestingly, the modification of the bands leads to an anisotropic quadratic degeneracy, as the curvature of the top band flips sign between orthogonal directions. These results show that the behaviour of the generalised Dirac cones discussed in this work is robust as long as the periodicity is kept small, but may change when the periodicity of the arrays is not very subwavelength $|\mathbf{a}_{1,2}|\gtrsim0.7\lambda_a$. Additionally, they show that accounting for fully retarded interactions in quantum metasurfaces is important, as their behaviour may strongly deviate from what a quasistatic approximation would predict, even giving rise to new effects such as the tilted Dirac cones discussed above. 
%

%

\section{Conclusions}

We discuss the presence and manipulation of generalised Dirac dispersions in a quantum
metasurface composed by single photon quantum emitters arranged in honeycomb arrays of subwavelength periodicity. We first reviewed the properties of Dirac cones and then we showed how modified dispersions emerge by introducing uniaxial anisotropy to the lattice, that is, by moving the two emitters in the unit cell farther or closer together along a given direction.
Specifically, we observe semi-Dirac points, which posses linear and quadratic dispersion in orthogonal directions, and tilted Dirac cones that change the local density of states at  
the degeneracy point form vanishing (type I) to diverging (type II and III). We explain the emergence and discuss the manipulation of each type of dispersion relation in the sub-radiant out-of-plane and in-plane modes, as well as their movement in reciprocal space with the anisotropy of the lattice. Moreover, we include a detailed discussion of the importance of including retardation effects to describe the system and how can affect the Dirac dispersions. Our results show that the behaviour of the generalised Dirac cones is robust to retardation effects, and that accounting for retarded interactions beyond quasistatic ones in the modelling of quantum metasurfaces is important in order to predict the correct behaviour of the modes.
The engineering and manipulation of such energy dispersions can modify substantially the quantum dynamics of local probes placed near the metasurface giving rise to new exotic forms of light-matter interactions.

\begin{acknowledgements}
 M.B.P. acknowledges support from the Basque Government's IKUR initiative on Quantum technologies (Department of Education). AGT acknowledges support from   CSIC Research   Platform   on   Quantum   Technologies PTI-001,  from  Spanish  project  PGC2018-094792-B-100(MCIU/AEI/FEDER, EU), and from the Proyecto Sin\'ergico CAM 2020 Y2020/TCS-6545 (NanoQuCo-CM). P.A.H. acknowledges funding from Funda\c c\~ao para a Ci\^encia e a Tecnologia and Instituto de Telecomunica\c c\~oes under projects UIDB/50008/2020, UTAPEXPL/NPN/0022/2021 and the CEEC Individual Program with reference CEECIND/02947/2020. 
\end{acknowledgements}

\bibliographystyle{apsrev4-1}
\bibliography{references} 

\end{document}